\documentclass[aps,showkeys,showpacs,superscriptaddress,a4paper,10pt]{revtex4-2}

\usepackage{amssymb}
\usepackage{amsmath}
\usepackage[dvips]{graphicx}
\usepackage[T2A]{fontenc}
\usepackage{graphicx}
\usepackage[breaklinks=true,colorlinks=true,linkcolor=blue,urlcolor=blue,citecolor=blue]{hyperref}
\usepackage{comment}
\DeclareMathOperator{\sign}{sign}

\begin{document}

\title[On the Modified Eguchi-Oki-Matsumura System]{On the Modified Eguchi-Oki-Matsumura System}

\author{P.~O.~Mchedlov-Petrosyan}
\author{L.~N.~Davydov}
\email[Corresponding author: ]{ldavydov@kipt.kharkov.ua}

\affiliation{National Science Center "Kharkiv Institute of Physics and Technology",   \\
1, Akademichna St., Kharkiv, 61108, Ukraine}

\begin{abstract}
To describe the simultaneous order-disorder transformation and phase separation Eguchi, Oki and Matsumura [\doi{10.1557/proc-21-589}] introduced the system of two equations: one equation, governing the evolution of a conserved order parameter, and the second equation for the non-conserved order parameter. The key feature of their model is the free energy functional, which contains the square gradient terms of the both order parameters and a fourth power polynomial depending on both order parameters. According to the general Hohenberg-Halperin classification it is the type C model. We show that if the dynamics of the conserved order parameter is governed by the convective-viscous Cahn-Hilliard equation, this system allows exact traveling wave solution.
\end{abstract}


\maketitle

\renewcommand{\theequation}{\arabic{section}.\arabic{equation}}
\section{Introduction}\label{s1}

In \cite{1}, to describe the simultaneous order-disorder transformation and phase separation, Eguchi, Oki and Matsumura introduced a system of two equations: one equation, governing the evolution of a conserved order parameter, and a second equation for the non-conserved order parameter. The key feature of their model is the free energy functional
\begin{equation} \label{1.1} F=\, \, \int \left[\frac{1}{2} \bar{\varepsilon }\left(\frac{\partial \bar{u}}{\partial x'} \right)^{2} +\frac{1}{2} \bar{\xi }\left(\frac{\partial \bar{v}}{\partial x'} \right)^{2} +f\left(\bar{u},\bar{v}\right)\, \right] dx'. \end{equation}
Here
\begin{equation} \label{1.2} f=\frac{1}{2} a\bar{u}^{2} +\frac{1}{4} r\bar{v}^{4} -\frac{1}{2} s\bar{v}^{2} +\frac{1}{2} g\bar{u}^{2} \bar{v}^{2} .  \end{equation}
In their continuous description the conserved parameter $\bar{u}$ was the local concentration and the non-conserved $\bar{v}$ -- the local degree of order The chemical potentials for $\bar{u}$ and $\bar{v}$ are defined as the variational derivatives $\bar{\mu }_{u} =\frac{\delta F}{\delta \bar{u}} ;\, \, \, \bar{\mu }_{v} =\frac{\delta F}{\delta \bar{v}} $:
\begin{equation} \label{1.3} \bar{\mu }_{u} =-\bar{\varepsilon }\frac{\partial ^{2} \bar{u}}{\partial x'^{2} } +\left(a+g\bar{v}^{2} \right)\bar{u}, \end{equation}
\begin{equation} \label{1.4} \bar{\mu }_{v} =-\xi \frac{\partial ^{2} \bar{v}}{\partial x'^{2} } +\left(g\bar{u}^{2} +r\bar{v}^{2} -s\right)\bar{v} .\end{equation}

The evolution of the conserved $\bar{u}$ and the non-conserved order parameter$\bar{v}$ is governed by the system of the Cahn-Hilliard (CH) equation \cite{2,3} and a nonlinear diffusion equation, usually called the Allen-Cahn (AC) equation \cite{4} in material science:
\begin{equation} \label{1.5} \frac{\partial \bar{u}}{\partial t'} =M\frac{\partial ^{2} \bar{\mu }_{u} }{\partial x'^{2} } =M\frac{\partial ^{2} }{\partial x'^{2} } \left[-\bar{\varepsilon }\frac{\partial ^{2} \bar{u}}{\partial x'^{2} } +\left(a+g\bar{v}^{2} \right)\bar{u}\right] ,\end{equation}
\begin{equation} \label{1.6} \frac{\partial \bar{v}}{\partial t'} =-K\bar{\mu }_{v} =K\left[\bar{\xi }\frac{\partial ^{2} \bar{v}}{\partial x'^{2} } -\left(g\bar{u}^{2} +r\bar{v}^{2} -s\right)\bar{v}\right]. \end{equation}
Here $M$ is the mobility and $K$ is a phenomenological coefficient. According to the general classification of Hohenberg and Halperin \cite{5} this is a type C model. In \cite{6} the connection between the discrete (lattice) approach and the continuum approach to the combination of conserved and non-conserved dynamics was studied in detail; after this work the name ``Cahn-Hilliard/Allen-Cahn system'' for these systems became commonly used. However, the behavior of such systems depends essentially on the form of the thermodynamic potential functional; so for definiteness we will call the system with the functional \eqref{1.1}-\eqref{1.2}  the Eguchi-Oki-Matsumura system (EOM-system). In \cite{1} two cases were considered: spinodal decomposition with ordering from a disordered state, and spinodal decomposition from a uniformly ordered state. The evolution of $\bar{u}$ and $\bar{v}$ is coupled through the constant $g$; to avoid confusion we point out that our notations are different from the notations of original papers. In \eqref{1.2} $a,\, r$ and $g$ are positive constants depending on the temperature, whereas $s$ increases from a negative to a positive value, as the temperature goes down from above to below the ordering temperature. The authors point out that the definition \eqref{1.1}-\eqref{1.2} while looking very specific in the form is actually quite general in the sense of fitting a wide class of phase diagrams. The model of \cite{1} was further studied mathematically and numerically, and applied to analyze the experiments in \cite{7,8,9,10,11,12,13,14,15,16}. There always finite domains, or mathematical initial-boundary value problems, were considered.

In our model we keep equation \eqref{1.6} for the evolution of the non-conserved order parameter; however the evolution of the conserved order parameter is governed by the convective-viscous Cahn-Hilliard (CH) equation:
\begin{equation} \label{1.7} \frac{\partial \bar{u}}{\partial t'} -2\bar{\alpha }\bar{u}\frac{\partial \bar{u}}{\partial x'} =M\frac{\partial ^{2} }{\partial x'^{2} } \left[-\bar{\varepsilon }\frac{\partial ^{2} \bar{u}}{\partial x'^{2} } +\left(a+g\bar{v}^{2} \right)\bar{u}+\bar{\eta }\frac{\partial \bar{u}}{\partial t'} \right] .\end{equation}
Here $\bar{\alpha }$ is proportional to the applied field, and $\bar{\eta }$ is the viscosity. Setting $\bar{\alpha }=0;\, \, \bar{\eta }=0$, i.e. taking the classic CH equation instead of the convective-viscous one, we arrive at the system \eqref{1.5}-\eqref{1.6}. We consider a one-dimensional infinite domain and look for the traveling wave solutions.

The classic Cahn-Hilliard equation was introduced as early as in 1958 \cite{2,3}; the stationary solutions were considered, the linearized version was treated, and corresponding instability of homogeneous state identified. However, an intensive study of the fully nonlinear form of this equation started essentially later \cite{17}. Now an impressive amount of work has been done on the nonlinear Cahn-Hilliard equation, as well as on its numerous modifications, see \cite{18,19}. An important modification was made by Novick-Cohen \cite{20}. Taking into account the dissipation effects which are neglected in the derivation of the classic Cahn-Hilliard equation, she introduced the \textit{viscous} Cahn-Hilliard (VCH) equation, with the term $\bar{\eta }\frac{\partial \bar{u}}{\partial t'} $ added to the chemical potential $\bar{\mu }_{u} $, see \eqref{1.7}. To account for the external field, several authors considered the nonlinear \textit{convective} Cahn-Hilliard equation (CCH) \cite{21,22,23}. To study the joint effects of nonlinear convection and viscosity, Witelski \cite{24} introduced the \textit{convective-viscous}-Cahn-Hilliard equation (CVCHE), see \eqref{1.7}, with a general symmetric double-well potential. With an additional constraint imposed on the nonlinearity and viscosity, approximate traveling-wave solutions were obtained. In \cite{25}, for the convective viscous CH equation several exact single- and two-wave solutions were obtained. Generally, the CH equation with convective and viscous terms appeared to be a useful tool in the study of further generalizations, such as external nonlinear sink/source terms and memory effects \cite{26,27}.

The present paper is organized as follows: in the next Section we find an exact traveling-wave solution for the system \eqref{1.6}-\eqref{1.7}. In Section 3 we consider the parametric dependence of the solution. In Section 4 we discuss our results.

\setcounter{equation}{0}
\section{Traveling wave solution}\label{s2}

Introducing the non-dimensional concentration $u=\frac{\bar{u}}{u_{0} } $, non-dimensional order parameter $v=\frac{\bar{v}}{v_{0} } $, non-dimensional coordinate $x=\frac{\bar{x}}{X} $ and non-dimensional time $t=\frac{t'}{T} $, we rewrite the system \eqref{1.7},\eqref{1.6} in non-dimensional form:
\begin{equation} \label{2.1} \frac{\partial u}{\partial t} -2\alpha u\frac{\partial u}{\partial x} =\frac{\partial ^{2} }{\partial x^{2} } \left[-\varepsilon \frac{\partial ^{2} u}{\partial x^{2} } +\left(a+v^{2} \right)u+\eta \frac{\partial u}{\partial t} \right], \end{equation}
\begin{equation} \label{2.2} \frac{\partial v}{\partial t} =D\frac{\partial ^{2} v}{\partial x^{2} } -\left(u^{2} +v^{2} -b\right)v ,\end{equation}
where it is convenient to take
\begin{equation} \label{2.3} T=\frac{g}{rK} ;\, \, \, \, X=\sqrt{TM} ;\, \, \, \, u_{0} =\frac{\sqrt{r} }{g} ;\, \, \, \, v_{0} =\frac{1}{\sqrt{g} } . \end{equation}
We also introduced the notations
\begin{equation} \label{2.4} \alpha =\frac{\bar{\alpha }}{\sqrt{gMK} } ;\, \, \, \, \, \varepsilon =\frac{\bar{\varepsilon }}{X^{2} } =\frac{\bar{\varepsilon }rK}{gM} ;\, \, \, \, b=s\frac{g}{r} ;\, \, \, \, D=\frac{K\xi }{M} . \end{equation}

Looking for the traveling wave solution, we introduce $z=x-\sigma t$; then Eqs.~\eqref{2.1}-\eqref{2.2} take the form
\begin{equation} \label{2.5} -\frac{d}{dz} \left(\sigma u+\alpha u^{2} \right)=\frac{d^{2} }{dz^{2} } \left[-\varepsilon \frac{d^{2} u}{dz^{2} } +\left(a+v^{2} \right)u-\sigma \eta \frac{du}{dz} \right] , \end{equation}
\begin{equation} \label{2.6} -\sigma \frac{dv}{dz} =D\frac{d^{2} v}{dz^{2} } -\left(u^{2} +v^{2} -b\right)v . \end{equation}
We look for special solutions of the system \eqref{2.5}-\eqref{2.6}, when the functional dependence $u=u\left(v\right)$ does not contain $z$ explicitly. Comparing the powers of $u,\, v\, $ in both equations we presume the proper Ansatz for the solution to be
\begin{equation} \label{2.7} u=\gamma v+\beta . \end{equation}
Substitution of the Ansatz into \eqref{2.5} and integration yields
\begin{eqnarray} \label{2.8}   &-&\alpha \gamma ^{2} \left[v^{2} +\frac{\sigma +2\alpha \beta }{\alpha \gamma } v+C\right] \nonumber \\ &=&\frac{d}{dz} \left[-\varepsilon \gamma \frac{d^{2} v}{dz^{2} } +\gamma v^{3} +\beta v^{2} +a\gamma v+\beta a-\sigma \gamma \eta \frac{dv}{dz} \right].  \end{eqnarray}
Here $C$ is the integration constant. Substitution of the Ansatz \eqref{2.7} into equation \eqref{2.6} yields
\begin{equation} \label{2.9} \sigma \frac{dv}{dz} =-D\frac{d^{2} v}{dz^{2} } +\left(\gamma ^{2} +1\right)\left[v^{2} +2\frac{\gamma \beta }{\left(\gamma ^{2} +1\right)} v+\frac{\beta ^{2} -b}{\left(\gamma ^{2} +1\right)} \right]v\, .\,  \end{equation}
Evidently, at $\pm \infty $, both the left-hand-side of \eqref{2.8} and the right-hand-side of \eqref{2.9} should be equal to zero; this means that the values of $v$ at $\pm \infty $ should be the common roots $v_{1,2} $ of the equations
\begin{equation} \label{2.10} v^{2} +\frac{2\alpha \beta +\sigma }{\alpha \gamma } v+C=0 ,\end{equation}
\begin{equation} \label{2.11} \left(v^{2} +\frac{2\gamma \beta }{\gamma ^{2} +1} v+\frac{\beta ^{2} -b}{\gamma ^{2} +1} \right)v=0 . \end{equation}

We are looking for a solution, which is non-zero at $\pm \infty $; then
\begin{equation} \label{2.12} C=\frac{\beta ^{2} -b}{\gamma ^{2} +1} ;\, \, \, \frac{2\alpha \beta +\sigma }{\alpha \gamma } =\frac{2\gamma \beta }{\gamma ^{2} +1} . \end{equation}
The second of these equalities yields an expression for the front velocity
\begin{equation} \label{2.13} \, \, \, \sigma =-\frac{2\alpha \beta }{\gamma ^{2} +1} . \end{equation}
The roots $v_{1,2} $ are
\begin{equation} \label{2.14} v_{1,2} =-\frac{1}{\gamma ^{2} +1} \left[\gamma \beta \pm \sqrt{b\left(\gamma ^{2} +1\right)-\beta ^{2} } \right]. \end{equation}

For the roots to be real, it should be
\begin{equation} \label{2.15} b\left(\gamma ^{2} +1\right)\ge \beta ^{2} . \end{equation}
So the necessary condition for the roots to be real is $b>0$; in \cite{1} this was the case of the possible ordering. For the solution approaching values $v_{1,2} $ at $\pm \infty $ the simplest possible Ansatz is
\begin{equation} \label{2.16} \frac{dv}{dz} =\kappa \left(v-v_{1} \right)\left(v-v_{2} \right)=\kappa \left(v^{2} -pv+q\right) .\end{equation}
Here, $\kappa >0$ for the anti-kink ($\frac{dv}{dz} <0$) and $\kappa <0$ for the kink ($\frac{dv}{dz} >0$).  We have denoted for brevity
\begin{equation} \label{2.17} p=-2\frac{\gamma \beta }{\left(\gamma ^{2} +1\right)} ;\, \, q=\frac{\beta ^{2} -b}{\left(\gamma ^{2} +1\right)} . \end{equation}
Then the polynomial in both \eqref{2.8} and\eqref{2.9} could be rewritten as
\begin{equation} \label{2.18} v^{2} +2\frac{\gamma \beta }{\left(\gamma ^{2} +1\right)} v+\frac{\beta ^{2} -b}{\left(\gamma ^{2} +1\right)} =\frac{1}{\kappa } \frac{dv}{dz} . \end{equation}
Substitution of \eqref{2.18} into \eqref{2.8}, \eqref{2.9} and integration yield
\begin{equation} \label{2.19} -\varepsilon \gamma \frac{d^{2} v}{dz^{2} } +\gamma v^{3} +\beta v^{2} +\left(a\gamma +\frac{\alpha \gamma ^{2} }{\kappa } \right)v+\beta a-\sigma \gamma \eta \frac{dv}{dz} +C_{1} =0 ,\end{equation}
\begin{equation} \label{2.20} -D\frac{dv}{dz} +\frac{\left(\gamma ^{2} +1\right)}{2\kappa } v^{2} -\sigma v+C_{2} =0\, . \,  \end{equation}
Using \eqref{2.16}, the second derivative is easily calculated
\begin{equation} \label{2.21} \frac{d^{2} v}{dz^{2} } =\kappa ^{2} \left[2v^{3} -3pv^{2} +\left(2q+p^{2} \right)v-pq\right] .\end{equation}
Substitution of the expressions for the derivatives transforms \eqref{2.19}-\eqref{2.20} into
\begin{eqnarray} \label{2.22} &&\left(1-2\varepsilon \kappa ^{2} \right)\gamma v^{3} +\left[3\varepsilon \gamma \kappa ^{2} p+\beta -\sigma \gamma \eta \kappa \right]v^{2} \nonumber  \\ &&+\left[-\varepsilon \gamma \kappa ^{2} \left(2q+p^{2} \right)+\gamma \left(a+\frac{\alpha \gamma }{\kappa } \right)+\sigma \gamma \eta \kappa p\right]v=0~,  \end{eqnarray}
\begin{equation} \label{2.23} \left[\frac{\left(\gamma ^{2} +1\right)}{2\kappa } -D\kappa \right]v^{2} +\left(D\kappa p-\sigma \right)v=0 .\end{equation}

In obtaining \eqref{2.22}-\eqref{2.23} we used the constants $C_{1} ,\, C_{2} $ to cancel $v$-independent terms. For $v$ defined by the Ansatz \eqref{2.16} to be simultaneously a solution of \eqref{2.8} and \eqref{2.9}, equations \eqref{2.22}-\eqref{2.23} should be satisfied identically for arbitrary $v$. That is, the coefficients of all powers of $v$ should be equal to zero. This yields five algebraic equations:
\begin{equation} \label{2.24} 1-2\varepsilon \kappa ^{2} =0 ,\end{equation}
\begin{equation} \label{2.25)} 3\varepsilon \kappa ^{2} \gamma p+\beta -\sigma \gamma \eta \kappa =0 , \end{equation}
\begin{equation} \label{2.26} -\varepsilon \kappa ^{2} \left(2q+p^{2} \right)+a+\frac{\alpha \gamma }{\kappa } +\sigma \eta \kappa p=0, \end{equation}
\begin{equation} \label{2.27} \left(\gamma ^{2} +1\right)-2D\kappa ^{2} =0 ,\end{equation}
\begin{equation} \label{2.28} D\kappa p-\sigma =0. \end{equation}

If the constraints \eqref{2.24}-\eqref{2.28} are satisfied, the solution of \eqref{2.16} is simultaneously a solution of \eqref{2.8} and \eqref{2.9}; integrating \eqref{2.16} and taking the position of the maximal value of the derivative $\left|\frac{dv}{dz} \right|$ as $z=0$, we get
\begin{equation} \label{2.29} v=\frac{v_{2} +v_{1} }{2} -\frac{v_{2} -v_{1} }{2} \tanh \left(\frac{1}{2} \kappa \left(v_{2} -v_{1} \right)z\right) .\end{equation}
Using \eqref{2.7} and \eqref{2.29}, we obtain the solution for $u$. While the form of the solutions is simple, the dependence of the parameters of the solutions on the system's parameters is complicated; it will be analyzed in the next Section.

\setcounter{equation}{0}
\section{The parametric dependence of the solution}\label{s3}

There are five constraints and only three unknowns, $\gamma ,\, \beta $, and $\kappa $. This means that for such a solution to exist two constraints should be imposed on the parameters of the system. Eliminating $\kappa ^{2} =\frac{1}{2\varepsilon } $ from the system of equations \eqref{2.24}-\eqref{2.28} (but keeping $\kappa $, as it may be positive or negative) yields
\begin{equation} \label{3.1} \frac{3}{2} \gamma p+\beta -\sigma \gamma \eta \kappa =0 ,\end{equation}
\begin{equation} \label{3.2} -q-\frac{1}{2} p^{2} +a+\frac{\alpha \gamma }{\kappa } +\sigma \eta \kappa p=0 , \end{equation}
\begin{equation} \label{3.3} \varepsilon \left(\gamma ^{2} +1\right)-D=0, \end{equation}
and Eq.~\eqref{2.28}, which we do not rewrite here. Multiplying \eqref{3.1} by $p$, \eqref{3.2} by $\gamma $ and summing, we get
\begin{equation} \label{3.4} -q\gamma +\gamma p^{2} +\beta p+\gamma a+\frac{\alpha \gamma ^{2} }{\kappa } =0 . \end{equation}
Substitution of $p,\, q$ and $\sigma $, given by \eqref{2.17} and \eqref{2.13} respectively, into \eqref{2.28}, \eqref{3.1} and \eqref{3.4} yields
\begin{equation} \label{3.5} \alpha =\gamma \kappa D ,\end{equation}
\begin{equation} \label{3.6} 2\gamma ^{2} =1+2\alpha \gamma \eta \kappa  ,\end{equation}
\begin{equation} \label{3.7} \left(3-\gamma ^{2} \right)\beta ^{2} =\left(\gamma ^{2} +1\right)\left[b+\left(\gamma ^{2} +1\right)\left(a+\frac{\alpha \gamma }{\kappa } \right)\right] . \end{equation}
Substitution of \eqref{3.5} for $\alpha $ into \eqref{3.7} yields
\begin{equation} \label{3.8} \beta ^{2} =\frac{\left(\gamma ^{2} +1\right)}{\left(3-\gamma ^{2} \right)} \left[b+\left(\gamma ^{2} +1\right)\left(a+\gamma ^{2} D\right)\right]\, . \end{equation}
From \eqref{3.8} it is evident, that
\begin{equation} \label{3.9} \gamma ^{2} <3 .\end{equation}

It is convenient to denote, see \eqref{3.3},
\begin{equation} \label{3.10} y=\gamma ^{2} +1=\frac{D}{\varepsilon } . \end{equation}
Then \eqref{3.8} can be rewritten as
\begin{equation} \label{3.11} \beta ^{2} =\frac{y}{\left(4-y\right)} \left[b+y\left(a-D+yD\right)\right]\, ; \end{equation}
and inequality \eqref{3.9} becomes $y<4$. The constraints \eqref{3.5} and \eqref{3.6} on the system parameters also impose a limitation on $y$. These constraints could be rewritten as
\begin{equation} \label{3.12} \, \, \gamma \kappa =\frac{\alpha }{D} ;\, \, \, \, \, \alpha \gamma \kappa \eta =y-\frac{3}{2} . \end{equation}
This means, that the product $\gamma \kappa $ should always have the same sign as the convective term $\alpha $; since the viscosity $\eta \ge 0$, it is necessary $y\ge \frac{3}{2} $. That is, together with \eqref{3.9}, the allowed values of $y$ are in the interval
\begin{equation} \label{3.13} \frac{3}{2} \le y<4 .\end{equation}
On the other hand, elimination of $\alpha $ from \eqref{3.5}-\eqref{3.6} yields
\begin{equation} \label{3.14} \eta =\frac{2y-3}{\left(y-1\right)y} .  \end{equation}

So the viscosity necessary for the existence of the solution is determined by $y$, i.e., by the ratio ${D\mathord{\left/ {\vphantom {D \varepsilon }} \right. \kern-\nulldelimiterspace} \varepsilon } $. Also, it is necessary to check the inequality \eqref{2.15}, i.e. the existence of the real $v_{1} ,\, v_{2} $,
\begin{equation} \label{3.15} b\ge \frac{1}{\left(3-\gamma ^{2} \right)} \left[b+\left(\gamma ^{2} +1\right)\left(a+\gamma ^{2} D\right)\right] , \end{equation}
or
\begin{equation}\label{3.16} b\ge \frac{1}{\left(4-y\right)} \left[y^{2} D+y\left(a-D\right)+b\right] . \end{equation}
It is shown in \textbf{{Appendix 1}} that for \eqref{3.16} to be satisfied, it must be
\begin{equation} \label{3.17} \, \, \, y_{1} <y<y_{2} . \end{equation}
Here
\begin{equation} \label{3.18} y_{1,2} =-\frac{\left(a+b-D\right)}{2D} \mp \frac{1}{2D} \sqrt{\left(a+b-D\right)^{2} +12bD} .  \end{equation}
However, the allowed interval should be compatible with \eqref{3.13}; this reduces it to
\begin{equation} \label{3.19} \frac{3}{2} \le y<y_{2} . \end{equation}
Additionally, this interval exists only if $\frac{3}{2} <y_{2} $, i.e., inequality \eqref{A1.6} is satisfied:
\begin{equation} \label{3.20} a+\frac{1}{2} D<b . \end{equation}
The expressions \eqref{3.3} and \eqref{3.8} determine the absolute values of $\gamma ,\, \beta $; however, there are additional physical limitations on their sign. Indeed, the concentration $u$ should be positive. So, if $v_{1} $ is the smaller root in \eqref{2.14}, it should be $\gamma v_{1} +\beta >0$:
\begin{equation} \label{3.21} \beta -\gamma \sqrt{b\left(\gamma ^{2} +1\right)-\beta ^{2} } >0 . \end{equation}

Evidently, if $\gamma >0;\, \, \beta <0$, the latter inequality is violated. If $\gamma >0$, it is necessary that $\beta >0$, and it should be $\, \beta ^{2} >\gamma ^{2} b$, or using notation \eqref{3.10},
\begin{equation} \label{3.22} \, \beta ^{2} >\left(y-1\right)b . \end{equation}
As it is shown in \textbf{{Appendix 2}}, the latter inequality is fulfilled if
\begin{equation} \label{3.23} y<\hat{y}_{2} ;\, \, \, or\, \, \, \hat{y}_{3} <y .  \end{equation}
Here $\hat{y}_{1} <\hat{y}_{2} <\hat{y}_{3} $ are the roots of the cubic polynomial \eqref{A2.2}. It is shown in \textbf{{Appendix 2}}, that if
\begin{equation} \label{3.24} \frac{15}{4} D+3a>b \end{equation}
$\hat{y}_{2} <\, \hat{y}_{3} <\frac{3}{2} $, so the interval \eqref{3.19} is not changed. On the other hand, if inequality \eqref{3.24} is violated, the allowed interval is split into two: $\left(\frac{3}{2} ,\, \hat{y}_{2} \right)$ and $\left(\hat{y}_{3} ,\, y_{2} \right)$. For the given numerical values, the roots of the cubic polynomial \eqref{A2.2} could be easily found; however, in the general form, the limits $\hat{y}_{2} ,\, \hat{y}_{3} $ are not informative. So we always presume the inequality \eqref{3.24} to be fulfilled, if $\beta >0,\, \, \gamma >0$.

If $\beta >0;\, \, \gamma <0$ the inequality \eqref{3.21} is evidently satisfied; so again, the allowed interval is\eqref{3.19}.
Finally, if $\beta <0;\, \, \gamma <0$, the inequality \eqref{3.21} is equivalent to
\begin{equation} \label{3.25} \, \beta ^{2} <\left(y-1\right)b .\end{equation}
Obviously, this inequality allows the presence of $y$ between the roots of the cubic polynomial \eqref{A2.2}.
\begin{equation} \label{3.26} \hat{y}_{2} <y<\hat{y}_{3} . \end{equation}
However, as it was discussed above, this is incompatible with $\frac{3}{2} >\hat{y}_{3} $; i.e. the combination $\beta <0;\, \, \gamma <0$ is impossible. The above considerations are summarized in Table 1.

\vspace{0.4cm}

Table 1
\begin{center}
\begin{tabular}{|p{1.1in}|p{2.0in}|p{1.8in}|} \hline
For $D+2a<2b$ & $\gamma >0$ & $\gamma <0$ \\ \hline
$\beta >0$  & $\frac{3}{2} \le y<y_{2} $; $\left(2b<\frac{15}{2} D+6a\right)$ & $\frac{3}{2} \le y<y_{2} $ \\ & $ \sign\left(\sigma \right)=-\sign\left(\alpha \right)$ & $\sign\left(\sigma \right)=-\sign\left(\alpha \right)$ \\ & $\sign\left(\kappa \right)=\sign\left(\alpha \right)$; & $\sign\left(\kappa \right)=-\sign\left(\alpha \right)$     \\ \hline
$\beta <0$  & impossible & impossible \\ \hline
\end{tabular}
\end{center}

The velocity sign selection is also shown in Table 1. So it is sufficient to consider the absolute value of the velocity; it is even more convenient to consider $\sigma ^{2} $ as a function of the parameters $y=\frac{D}{\varepsilon } $, $a,\, b,$ and $D$:
\begin{equation} \label{3.27}  \sigma ^{2} =2D\frac{\left(y-1\right)}{\left(4-y\right)} \left[Dy^{2} +y\left(a-D\right)+b\right] .\end{equation}
So the velocity is non-zero in the allowed interval of $y$.

\setcounter{equation}{0}
\section{Discussion}\label{s4}

In the present paper, we obtained an exact traveling wave solution for the modified Eguchi-Oki-Matsumura model \cite{1}. The EOM-system describes the combined dynamics of conserved and non-conserved order parameters. While in the original EOM system the conserved dynamics was governed by the classic CH equation, we have used the convective-viscous CH equation, introduced in \cite{23}. We looked for special solutions, when the functional link between conserved $u$ and non-conserved $v$ order parameters is independent of the traveling wave coordinate $z$. Both waves travel with the same velocity. For such link to exist, additional constraints should be imposed on the system parameters. We have found that the conserved order parameter (concentration) should be a linear function of the non-conserved order parameter. The existence of such a linear function and a physically reasonable choice of the signs of its coefficients imposes limitations on the parameter $y=\frac{D}{\varepsilon } $. It is practical to rewrite this parameter in the initial dimensional variables
\begin{equation} \label{4.1} y=\frac{D}{\varepsilon } =\frac{\xi g}{\bar{\varepsilon }r} . \end{equation}

$\bar{\varepsilon }$ is the coefficient at the $\left(\frac{\partial \bar{u}}{\partial x'} \right)^{2} $term and $\xi $ is the coefficient at the $\left(\frac{\partial \bar{v}}{\partial x'} \right)^{2} $ term in the free energy functional. The constant $g$ represents the mutual influence of the conserved and non-conserved order parameters; i.e., $y$ reflects evidently the `degree of coupling' of the conserved and non-conserved dynamics. The lower and upper limits in the necessary constraint $\frac{3}{2} \le y<y_{2} $ are of a different nature. The lower limit, $y=\frac{3}{2} $, corresponds to the special case of zero viscosity; the conserved dynamics is without dissipation. The upper limit corresponds to a zero discriminant in \eqref{2.14}, $v_{1} =v_{2} $, i.e., the amplitude of the waves of both order parameters collapses to zero. The coefficients $\beta ,\, \, \gamma $, see \eqref{2.7}, remain finite for all allowed $y$; while $\beta >0$, $\gamma $ could be positive or negative; however, it is always $sign\left(\gamma \kappa \right)=sign\left(\alpha \right)$. The sign of the velocity is always opposite to the sign of the convective term. The possible kink/anti-kink combinations for $u$ and $v$ solutions are shown in the Table 2

\vspace{0.4cm}

Table 2
\begin{center}
\begin{tabular}{|p{1.5in}|p{1.5in}|p{1.5in}|} \hline
 & $\kappa >0$  & $\kappa <0$  \\ \hline
$\alpha >0$, \,\,\,  $\sigma <0$  & $\gamma >0$ \newline $v$--anti kink; $u$--anti kink   & $\gamma <0$\newline  $v$--kink; $u$--anti kink \\ \hline
$\alpha <0$, \,\,\,  $\sigma >0$  & $\gamma <0$ \newline $v$--anti kink; $u$--kink & $\gamma >0$ \newline $v$--kink; $u$--kink \\ \hline
\end{tabular}
\end{center}

\noindent For the existence of the allowed interval $\left[\frac{3}{2} ,\, y_{2} \right]$, it is necessary that
\begin{equation} \label{4.2} a+\frac{1}{2} D<b. \end{equation}
In the dimensional parameters, the latter inequality is
\begin{equation} \label{4.3} a+\frac{K\xi }{2M} <s\frac{g}{r} .  \end{equation}

It was pointed out in \cite{1} that $s>0$ is necessary for the ordering. Inequality \eqref{4.3} shows that for the coupled waves of ordering and concentration change to exist the product of $s$ and coupling constant $g$ should be large enough. Additionally, when both $\beta ,\, \gamma $ are positive, to keep the same allowed interval for $\gamma ^{2} +1$, the parameter $b$ should be also limited from above, see \eqref{3.24}.

As it was already mentioned, if $y>\frac{3}{2} $, for the adjustment of the two processes the non-zero viscosity is needed. Viscosity $\eta $ as a function of $y$, see \eqref{3.14}, has a maximum at $\, \bar{y}_{2} =\frac{3}{2} +\frac{\sqrt{3} }{2} $, see \textbf{Appendix 3}. However, for $\bar{y}_{2} <y_{2} $, i.e., to be in the allowed interval of $y$, the parameter $b$ should be larger, than that given by \eqref{4.2}:
\begin{equation} \label{4.4} \left(2+\sqrt{3} \right)\left[\, \left(1+\sqrt{3} \right)D+2a\right]<2b. \end{equation}
If the latter inequality is violated, the viscosity $\eta \left(y\right)$ necessary for the existence of solutions increases monotonically in the allowed interval of $y$.

The value of $D$ expressed in the initial dimensional parameters equals  $\frac{K\xi }{M} $, i.e., is proportional to the ratio of the kinetic coefficient for the non-conserved order parameter and the mobility. The velocity \eqref{3.27} increases monotonically with $D$, i.e., the faster the non-conserved kinetics, the higher the velocity of the waves. Of course, $D$ is limited by the inequality \eqref{4.2}.

\setcounter{equation}{0}
\subsection*{Appendix 1} \label{Sec:A1}
\renewcommand{\theequation}{A1.\arabic{equation}}


Here we check the inequality \eqref{3.16}, keeping in mind $\frac{3}{2} \le y<4$:
\begin{equation} \label{A1.1} b\ge \frac{1}{\left(4-y\right)} \left[y^{2} D+y\left(a-D\right)+b\right] . \end{equation}
The equivalent inequality is
\begin{equation} \label{A1.2} P_{2} \left(y\right)=y^{2} +y\frac{a+b-D}{D} -\frac{3b}{D} \le 0 . \end{equation}
For this inequality to be satisfied, it should be $y_{1} <y<y_{2} $, where $y_{1,2} $ are the roots of the polynomial $P_{2} \left(y\right)$:
\begin{equation} \label{A1.3} y_{1,2} =-\frac{1}{2D} \left[\left(a+b-D\right)\pm \sqrt{\left(a+b-D\right)^{2} +12bD} \right]. \end{equation}
Evidently, the real roots exist; we still need to check, whether $\frac{3}{2} \le y<4$ and $y_{1} <y<y_{2} $ are compatible. It appears that $y_{2} <4$ :
\begin{equation} \label{A1.4} \sqrt{\left(a+b-D\right)^{2} +12bD} -\left(a+b-D\right)<8D . \end{equation}
The latter inequality is equivalent to
\begin{equation} \label{A1.5} 0<4a+b+12D .\end{equation}

Inequality \eqref{A1.5} is evidently satisfied. Now we need to check, whether $\frac{3}{2} <y_{2} $; however, it is possible only if
\begin{equation} \label{A1.6} a+\frac{1}{2} D<b . \end{equation}
Additionally, comparing $y_{1} $ and $\frac{3}{2} $ we see that $y_{1} <\frac{3}{2} $ is equivalent to the evidently correct inequality:
\begin{equation} \label{A1.7} -\sqrt{\left(a+b-D\right)^{2} +12bD} <a+b+2D .\end{equation}
Finally, the allowed interval is reduced to
\begin{equation} \label{A1.8} \frac{3}{2} \le y<y_{2} . \end{equation}
This interval exists only if \eqref{A1.6} is satisfied.

\setcounter{equation}{0}
\subsection*{Appendix 2} \label{Sec:A2}
\renewcommand{\theequation}{A2.\arabic{equation}}


Here we check the inequality \eqref{3.22}, keeping in mind that $\frac{3}{2} \le y<4$
\begin{equation} \label{A2.1} \frac{y}{\left(4-y\right)} \left[Dy^{2} +y\left(a-D\right)+b\right]>\left(y-1\right)b .\end{equation}
The latter inequality is equivalent to
\begin{equation} \label{A2.2} P_{3} \left(y\right)=y^{3} +\frac{a+b-D}{D} y^{2} -\frac{4b}{D} y+\frac{4b}{D} >0 .\end{equation}
Looking for the extrema of $P_{3} \left(y\right)$, we get
\begin{equation} \label{A2.3} \tilde{y}^{2} +2\frac{a+b-D}{3D} \tilde{y}-\frac{4b}{3D} =0 ,\end{equation}
\begin{equation} \label{A2.4} \tilde{y}_{1,2} =-\frac{a+b-D}{3D} \mp \frac{1}{3D} \sqrt{\left(a+b-D\right)^{2} +12bD} . \end{equation}
Comparing the root $\tilde{y}_{2} $ with the root $y_{2} $ given by \eqref{A1.3} shows that $\tilde{y}_{2} <y_{2} $ because:
\begin{equation} \label{A2.5}  a+b-D<\sqrt{\left(a+b-D\right)^{2} +12bD} . \end{equation}

This means that at $y_{2} $, there is a rising part of the cubic curve. On the other hand, multiplying $P_{2} \left(y\right)$, see  \eqref{A1.2}, by $y$ and subtracting from $P_{3} \left(y\right)$, see \eqref{A2.2}, we get
\begin{equation} \label{A2.6} \delta \left(y\right)=P_{3} \left(y\right)-yP_{2} \left(y\right)=\frac{b}{D} \left(4-y\right) .\end{equation}
Now, $\delta \left(y_{1,2} \right)=P_{3} \left(y_{1,2} \right)-yP_{2} \left(y_{1,2} \right)=P_{3} \left(y_{1,2} \right)$ :
\begin{equation} \label{A2.7} \frac{D}{b} P_{3} \left(y_{1,2} \right)=\left(4-y_{1.2} \right)=4+\frac{1}{2D} \left[\left(a+b-D\right)\pm \sqrt{\left(a+b-D\right)^{2} +12bD} \right] .\end{equation}
It is easy to show, that $P_{3} \left(y_{2} \right)>0$; so $P_{3} \left(y_{1} \right)>P_{3} \left(y_{2} \right)>0$ too.
This means, that the largest root $\, \hat{y}_{3} $ of the cubic equation $P_{3} \left(y\right)=0$ is smaller than $y_{2} $; $y_{1} <y_{2} $ but $P_{3} \left(y_{1} \right)>P_{3} \left(y_{2} \right)$, so at $y_{1} $ the cubic curve is descending. That is, the two larger roots $\, \hat{y}_{2} \, ,\, \hat{y}_{3} $ of the cubic equation are within the interval $\left(y_{1} ,\, y_{2} \right)$. If $D+2a<2b$, then $\frac{3}{2} <y_{2} $. Calculation of $P_{3} \left(y\right)$ at $y=\frac{3}{2} $ yields
\begin{equation} \label{A2.8} \left. P_{3} \left(y\right)\right|_{y=\frac{3}{2} } =\frac{1}{8D} \left[9D+18a+2b\right]>0 .\end{equation}
So the point $\frac{3}{2} $ is outside the interval $\left(\hat{y}_{2} ,\, \hat{y}_{3} \right)$; calculation of $\frac{dP_{3} \left(y\right)}{dy} $ at $y=\frac{3}{2} $ yields
\begin{equation} \label{A2.9} \left. \frac{dP_{3} \left(y\right)}{dy} \right|_{y=\frac{3}{2} } =\frac{1}{D} \left(\frac{15}{4} D+3a-b\right) .\end{equation}

Together with \eqref{A2.8}, his means, that if
\begin{equation} \label{A2.10} \frac{15}{4} D+3a-b>0 \end{equation}
at $y=\frac{3}{2} $ the curve is rising, and $\hat{y}_{3} <\frac{3}{2} $. Then the allowed interval of $y$, see \eqref{3.17}, is
\begin{equation} \label{A2.11} \frac{3}{2} \le y<y_{2}  ,  \end{equation}
which coincides with \eqref{A1.7}. On the other hand, if \eqref{A2.10} is violated, the allowed interval is `split' into two parts: $\left(\frac{3}{2} ,\, \hat{y}_{2} \right)$ and $\left(\hat{y}_{3} ,\, y_{2} \right)$.

\setcounter{equation}{0}
\subsection*{Appendix 3} \label{Sec:A3}
\renewcommand{\theequation}{A3.\arabic{equation}}


Looking for the extrema of $\eta \left(y\right)$, see \eqref{3.14}, yields
\begin{equation} \label{A3.1}  \, \bar{y}_{1,2} =\frac{3}{2} \mp \frac{\sqrt{3} }{2} . \end{equation}
Comparing $\, \bar{y}_{2} $ with $y_{2} $ given by \eqref{A1.3} shows that $\bar{y}_{2} <y_{2} $ only if
\begin{equation} \label{A3.2} \left(2+\sqrt{3} \right)\left[\, \left(1+\sqrt{3} \right)D+2a\right]<2b. \end{equation}
The latter inequality for $b$ is evidently stronger, than \eqref{A1.6}. It means that for $\eta \left(y\right)$ to have a maximum in the allowed interval of $y$ the parameter $b$ should be essentially larger, than that in \eqref{A1.6}.

{\small \topsep 0.6ex

}

\end{document}